\documentclass[a4paper,12pt,doublespace]{article}

\usepackage{subfig}

\usepackage{moreverb}
\usepackage{appendix}
\usepackage{multirow}   

\usepackage[pdftex,colorlinks,bookmarksopen,bookmarksnumbered,citecolor=red,urlcolor=red]{hyperref}

\usepackage{amsmath}
\usepackage{multicol}   

\usepackage{dcolumn}

\newcolumntype{.}{D{.}{.}{4}}

\usepackage{setspace}
\usepackage[left=2cm,top=2cm,right=2cm,bottom=2cm]{geometry}

\usepackage{rotating}
\usepackage{graphicx}
\usepackage{color}
\usepackage{amsmath}
\usepackage{amsfonts}
\usepackage{amssymb}
\usepackage{soul}

\usepackage[numbers, sort&compress]{natbib}
 \makeatletter 
 \renewcommand\@biblabel[1]{#1.} 
 \makeatother %

\begin{document}


\centering
\Large{Bayesian Hierarchical Modeling of Longitudinal Glaucomatous Visual Fields using a 
Two-Stage Approach}

\vspace{10 mm}

\centering
\large{S. R. Bryan$^{a,b}$, P.H.C. Eilers$^{a}$, B. Li$^{a}$, D. Rizopoulos$^{a}$, K.A. Vermeer$^{b}$, \\  H.G. Lemij$^{c}$  and E.M.E.H. Lesaffre$^{a,d}$}

\vspace{10 mm}

\normalsize{$a$ Department of Biostatistics, Erasmus MC, Rotterdam, the Netherlands\\
         $b$ Rotterdam Ophthalmic Institute, Rotterdam, The Netherlands\\
         $c$ Glaucoma Service, Rotterdam Eye Hospital, Rotterdam, The Netherlands\\
         $d$ L-Biostat, KU Leuven, Leuven, Belgium}

\vspace{10 mm}

\normalsize{Corresponding author: Susan Bryan, Department of Biostatistics, Erasmus MC, PO Box 2040, 3000 CA Rotterdam, The Netherlands\\
email: s.bryan@eramsusmc.nl, Tel: +31/(0)10/7044231, Fax: +31/(0)10/7043014}

\singlespacing
\abstract{The Bayesian approach has become increasingly popular because it allows to model quite complex models via Markov chain Monte Carlo (MCMC) sampling. However, it is also recognized nowadays that MCMC sampling can become computationally prohibitive when a complex model needs to be fit to a large data set. To overcome this problem, we applied and extended a recently proposed two-stage approach to model a complex hierarchical data structure of glaucoma patients who participate in an ongoing Dutch study. Glaucoma is one of the leading causes of blindness in the world. In order to detect deterioration at an early stage, a model for predicting visual fields (VF) in time is needed. Hence, the true underlying VF progression can be determined, and treatment strategies can then be optimized to prevent further VF loss. Since we were unable to fit these data with the classical one-stage approach upon which the current popular Bayesian software is based, we made use of the two-stage Bayesian approach. The considered hierarchical longitudinal model involves estimating a large number of random effects and deals with censoring and high measurement variability. In addition, we extended the approach with tools for model evaluation}

\newpage

\doublespacing

\section{Introduction}

Since the introduction of MCMC sampling by Gelfand and Smith \cite{gelfand;smith;90} and the development of the BUGS software \cite{Lunn_2009} the Bayesian approach has become tremendously popular in various application areas, but especially to fit models to complex data structures. But with the years it also became clear that MCMC sampling can be computationally quite cumbersome, and even prohibitive, for fitting complex models to relatively large data sets. Several attempts have been made to look for alternative computational procedures and software, with notable examples such as INLA \cite{Rue_2009} and STAN \cite{Hoffman_2014}. While this newly software can sometimes  speed up the computations considerably, the computational gain is not always obvious upfront and for some advanced models the new developments may not be suitable yet. In addition, the majority of the practical Bayesians still use BUGS-related software. In this context, Lunn et al. \cite{Lunn_2013} proposed to fit a hierarchical model in two stages. The authors claim more model flexibility in this way, but advocate the use of their procedure especially for its computational properties. In this paper we further illustrate the use of the two-stage approach on a far more complex hierarchical data structure of glaucoma patients. In addition, we extended the approach with an additional sampling step to allow for the calculation of model selection and model evaluation criteria.

Our modeling approach is motivated by data from the Glaucoma Study conducted by the Rotterdam Eye Hospital in the Netherlands. According to the World Health Organization (WHO), glaucoma is one of the leading causes of irreversible blindness in the world \cite{kingman_2004}. Adequate treatment may slow down the disease, possibly even halting its progression. Evaluation of a longitudinal series of visual fields (VF), as measured by standard automated perimetry (SAP), provides a way to detect early evidence of glaucoma and to determine functional deterioration. However, due to the subjective nature of this technique, SAP is prone to large variability. In order to measure the true progression of the disease, this variability needs to be taken into account. The Glaucoma Study provides a unique database with a long follow up time. Although methods may have existed to model this type of data, the difficulties in extracting it from the device has made this type of data rare and hence has prevented much research on the topic.

The response variable of interest are the sensitivity estimates which describe the level of differential light sensitivity at different locations within each eye. The sensitivity estimates are left-censored due to the limitation of the device. Models which take into account this type of censoring, such as the Tobit model, have been described in the literature \cite{tobin_1958}. Our interest lies in modeling the latent, true values rather than the observed sensitivity estimates for two reasons. Firstly, clinical interest lies in predicting the disease progression rather than the observed sensitivity estimates. Secondly, using the latent scale allows us to use a simpler model than when directly modeling the observed data. The hierarchical structure of the data consists of 4 levels, namely, (1) the individual, (2) the eye (3) the hemifield and (4) the location. There is a vast amount of literature that addresses hierarchical mixed effects models, for both frequentist \cite{verbeke_2009} and Bayesian \cite{ntzoufras_2009, lesaffre_2012} approaches. We model this complex data structure using a Bayesian hierarchical mixed effects model with cross-classified random effects. Hence, we combine both spatial and time effects. One of the difficulties in modeling VF data is the amount and type of measurement error or variability in the sensitivity estimates. This may be due to measurable factors, such as season, time of day and reliability indices, or unknown transient factors, such as fatigue, lack of concentration, or delayed reaction time. Although their magnitudes may vary, these factors affect all locations belonging to the same VF. We propose to  model them as Global Visit Effects (GVEs). Furthermore, there is an inverse relationship between sensitivity and variability. For example, measurement error in the VFs increases with damage, and hence low sensitivity estimates have high variability. Therefore, it is naive to assume a constant variance over the wide range of sensitivity estimates. In this paper, we relax this assumption in order to incorporate this relationship. A problem with high dimensional data and complex data structures, is that it is sometimes difficult or even impossible to model them with standard MCMC algorithms. Lunn et al. \cite{Lunn_2013} proposed a two-stage approach, which allowed us to simplify the problem while still benefiting from the advantages of a full Bayesian model. However, one of the disadvantages of this approach, is that it is not possible to directly obtain the random effects estimates needed for most model evaluations. We address this issue by extending the two-stage approach to be able to determine these estimates.

Our aim is to model this complex data structure in order to obtain better estimates of the true evolution of the sensitivity over time, so that treatment strategies can be optimized to prevent further progression of VF loss. The structure of the paper is as follows. In Section \ref{GlaucomaStudy} we give further details on the motivating data set and introduce the research questions that triggered our modeling approach(es). In Section \ref{methods} we describe the models used in the analysis. In the subsequent section we briefly review computational aspects of the analysis. Model comparison is dealt with in Section \ref{evaluation}. In Section \ref{Application} we apply our models to the Glaucoma Study data. Section \ref{Discussion} contains a concluding discussion. Further details regarding the modeling approach are provided in an appendix.

\section{Motivating data set: the Glaucoma Study} \label{GlaucomaStudy}

\subsection{Description of the project} \label{GlaucomaStudy_description}

The Glaucoma Study is a prospective cohort study conducted by the Rotterdam Eye Hospital in the Netherlands. This is an ongoing study which began in 1998. Inclusion criteria included glaucoma diagnosis and an age range of 18 to 85 years. In total, 139 patients, consisting of 80 (57.6\%) men and 59 (42.4\%) women, were recruited with a mean follow-up of 10.5 years. Follow-up data were collected at approximately 6-monthly intervals. All patients gave their written informed consent for participation. All research procedures followed the tenets set forth in the Declaration of Helsinki. Furthermore, all of the data that was used in this analysis has been made available online at http://rod-rep.com.

\begin{figure}[ht]
\begin{center}
\includegraphics[width=.45\textwidth]{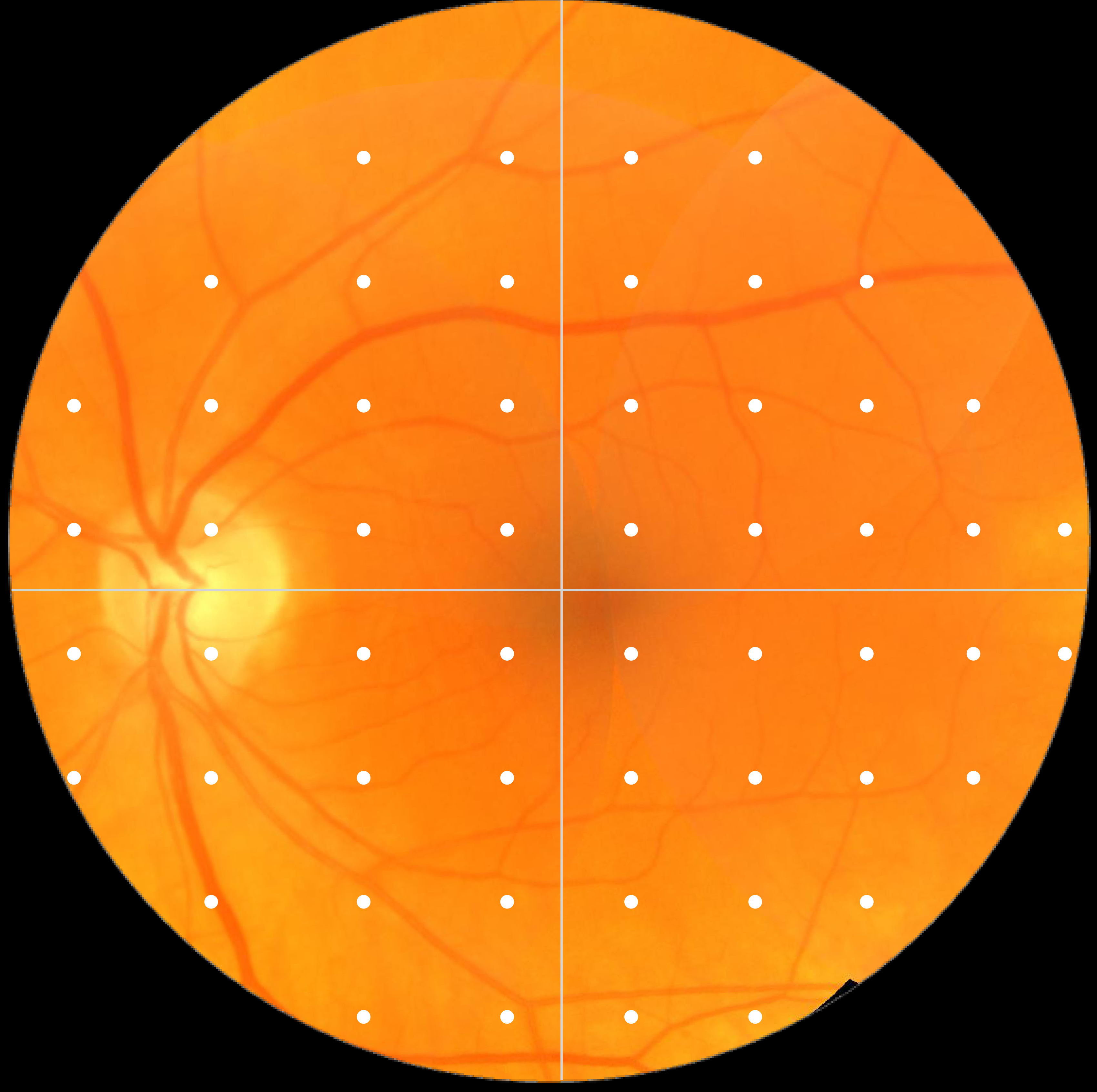}
\caption{Fundus photo of a left human eye with the 54 test locations for the VF test represented by white dots.}
\label{fig_fundus}
\end{center}
\end{figure}

Sensitivity estimates were measured at 52 test locations within each eye, or 26 test locations within each hemifield (excluding two locations corresponding to the blind spot) as shown in Figure \ref{fig_fundus}. The VFs were tested using the Humphrey Field Analyzer with the 24-2, white-on-white test strategy using the Full Threshold algorithm. The light source can be attenuated in the range from 1 to 10,000 times. On the decibel (dB) scale an attenuation $x$ is defined as $s = 10 \mbox{ \tt{log}}_{10}(x)$, or $x = 10^{s / 10}$. The lowest sensitivity that can be detected by this perimeter is 0 dB, although negative values could in fact occur if it were not for the limitations of this device. The highest sensitivity that can be detected is 50 dB, however few humans are capable of seeing a stimulus less than 40 dB, which is 1/10,000 of the maximum intensity of the instrument (or 1 asb). Thus, for practical purposes, the useful intensity range for white light testing is from 0 to 40 dB with a background illumination of 31.5 asb.  \cite{anderson_1999}.

\subsection{Previous research} \label{prev_research}

Parameters such as the mean deviation (MD) and visual field index (VFI) summarize the 52 sensitivity estimates into single values which can be used by the clinicians when optimizing treatment strategies. Longitudinal modeling of these VF summary parameters has been done before \cite{bengtsson_2008,artes_2011,kymes_2012,montolio_2012}. Modeling of individual test locations is potentially of greater interest, because it provides additional information such as the spatial nature of the fields which is otherwise lost in global parameters. In previous research, each location was analyzed as an independent sample \cite{mcnaught_1995, caprioli_2011, Bryan_2013}. However, separate location-specific regression models are not able to use any information from the data set as a whole. Multilevel mixed-effects models provide a better fit to the data than separate regression models by accounting for group effects and/or within-group correlation \cite{verbeke_2009}. This was shown in the context of global VF measurements by Pathak et al. \cite{pathak_2013}.

In glaucoma, variability is presumably related to fatigue effects and response errors, whereby sensitivity estimates decrease over time \cite{hudson_1994, Bengtsson_2000}. Differences in fatigue effects, between the inferior and superior hemifields within an eye have been demonstrated \cite{hudson_1994}.  Furthermore, this effect may differ between the first and second eye at the same visit. The number of false-negative answers have been shown to be higher in eyes with field loss. This may be explained by an increased variability in sensitivity estimates typically found in such eyes \cite{Bengtsson_2000, Russell_2012}. A common approach to reduce measurement variability is to average multiple measurements. For example, random uncorrelated measurement errors that are present in the point-wise sensitivity estimates are reduced when calculating summary parameters such as the mean deviation (MD). Other errors, however, are spatially correlated and affect the whole VF. One group of such errors are measurable factors, including season, time of day and reliability indices, which have been evaluated before \cite{montolio_2012}. Although these factors are statistically significant, they are rather small and hence only explain a small part of the observed global variation in VFs.

The inverse relationship between variability and sensitivity has been described in the literature. Henson et al. (2000) \cite{Henson_2000} found that this relationship is well represented by the function $\log(SD)=A + B \times {\tt sensitivity} (dB)$, where A and B are 2.81 dB and -0.066 dB respectively for normal eyes and
3.62 dB and -0.098 dB for glaucomatous eyes. Russell et al. (2012) \cite{Russell_2012} showed that the distribution of residuals is relatively concentrated at high VF sensitivities (26 to 36 dB) but stretches substantially as the sensitivity estimates decrease to a level of 10 dB. Sensitivity estimates near 10 dB are associated with residuals spanning almost the entire dynamic range of the instrument. This could be caused by a loss of ganglion cells (due to glaucomatous damage), or relocation of the stimulus to the peripheral visual field where there are fewer ganglion cells \cite{swanson_2011}. Zhu et al. (2014) \cite{Zhu_2014} describe a method to detect change using an inferential statistical model which incorporates the non-stationary variability using a mixture of Weibull distributions.

Although there is a wide range of literature which discusses these aspects, the majority of previous work deals with the global indices or treats each point-wise estimate as an independent sample. Furthermore, these aspects have been addressed separately. Hence, it is clear that an approach which takes into account the complex structure of the data and considers all of the aforementioned problems, is needed. We will address censoring, the hierarchical structure, the global variation as well as the relationship between the variability and sensitivity.

\section{Statistical Models} \label{methods}

Modeling the sensitivity estimates is beneficial for the evaluation of the progression of VF loss. By incorporating biological effects into the model, we aimed to improve the model fit and hence provide a better method for modeling this progression. This was done by building the model up sequentially.

\subsection{Censoring}\label{censoring}

It is important to note that unseen sensitivity estimates are indicated on the VF print out as $< 0$. They are smaller than zero because the instrument is unable to determine such sensitivities. Thus a model which defines the relationship between time and the latent, true sensitivity value is needed. The relationship between the observed $y^*$  and the latent, true sensitivity value $y$ is given by,

\begin{eqnarray*}
y^*  = y  \times I(y \geq 0) \mbox{ } + \mbox{ } 0 \times I(y < 0).
\end{eqnarray*}

\subsection{Hierarchical Model}\label{hierarchical}
We propose using a Bayesian hierarchical mixed effects model \cite{ntzoufras_2009, lesaffre_2012} to analyze the glaucoma data. This model is able to take into account both the within subject and between subject variability. Furthermore, we capitalized on the common features within each eye by taking into account the correlation between measurements belonging to the same eye. In addition, correlation of VF measurements within the inferior and superior hemifields, separated by the horizontal raphe, was assumed to be higher than between hemifields. Hence, the hierarchical structure of the data consists of 4 levels, namely, (1) the individual, (2) the eye (3) the hemifield and (4) the location. Let $\beta$ correspond to the regression parameters and $\mbox{\tt{years}}_{it}$ represent the time between measurement $t$ and the first measurement for each individual $i$, ranging from 0 to 10.5 years. The individual-specific intercept and slope are represented by $\alpha$, the eye-specific intercept and slope by $\gamma$, the hemifield-specific intercept and slope by $\eta$, and the location-specific intercept and slope by $\lambda$. We then have, for individual $i = 1, \dots , N$; eye $e = 1, 2$; hemifield $h=1,2$; location $l=1, \dots , 26$ and timepoint $t=1, \dots , T_i$,
\newline

Model 1:
\begin{eqnarray}\label{both_eyes}
y_{iehlt} &=& \beta_0 + \beta_1 \mbox{\tt{years}}_{it} +  \alpha_{0i} + \alpha_{1i} \mbox{\tt{years}}_{it} + \gamma_{0ie} + \gamma_{1ie} \mbox{\tt{years}}_{it}  + \eta_{0ieh} + \eta_{1ieh} \mbox{\tt{years}}_{it}+\nonumber \\ && \lambda_{0iehl} + 
 \lambda_{1iehl} \mbox{\tt{years}}_{it} + \epsilon_{iehlt} \\
&=& \mu^{(1)}_{iehlt} + \epsilon_{iehlt}, \nonumber
 \end{eqnarray}

where \begin{eqnarray*}
  \alpha_{i}  &=&  \left( \begin{smallmatrix} \alpha_{0i} \\ \alpha_{1i} \end{smallmatrix} \right) \sim
  				  N(\left( \begin{smallmatrix} 0 \\ 0 \end{smallmatrix} \right) ,
  				  \Sigma_\alpha = \left(\begin{smallmatrix} \Sigma_{\alpha11} &\Sigma_{\alpha12}  \\ \Sigma_{\alpha21} & \Sigma_{\alpha22} 		  							  \end{smallmatrix} \right) ) ; \\
 \gamma_{ie}  &=& \left( \begin{smallmatrix} \gamma_{0ie} \\ \gamma_{1ie} \end{smallmatrix} \right) \sim
  				  N(\left( \begin{smallmatrix} 0 \\ 0 \end{smallmatrix} \right) , \Sigma_\gamma = (\begin{smallmatrix} 						  						  \Sigma_{\gamma11} &\Sigma_{\gamma12}  \\ \Sigma_{\gamma21} & \Sigma_{\gamma22} \end{smallmatrix} )); \\
 \eta_{ieh}  &=&  \left( \begin{smallmatrix} \eta_{0ieh} \\ \eta_{1ieh} \end{smallmatrix} \right) \sim
  					N (( \begin{smallmatrix} 0 \\ 0 \end{smallmatrix}) ,
  					\Sigma_\eta = (\begin{smallmatrix} \Sigma_{\eta11} &\Sigma_{\eta12}  \\ \Sigma_{\eta21} & \Sigma_{\eta22} 	
  					\end{smallmatrix}) )  ; \\
 \lambda_{iehl} &=&  \left( \begin{smallmatrix} \lambda_{0iehl} \\ \lambda_{1iehl} \end{smallmatrix} \right) \sim
 					N ( \left( \begin{smallmatrix} 0 \\ 0 \end{smallmatrix} \right) ,
 					\Sigma_\lambda = ( \begin{smallmatrix} \Sigma_{\lambda11} &\Sigma_{\lambda12}  \\ \Sigma_{\lambda21} & \Sigma_{\lambda22} 	
 													\end{smallmatrix}))  \mbox{ } \mbox{and} \mbox{ }\\
 					\epsilon_{iehlt} &\sim& N(0,\sigma^2) .
\end{eqnarray*}

\subsection{Visit Effect}\label{visit_effect}

\begin{figure}[ht]
\begin{center}
\includegraphics[width=.7\textwidth]{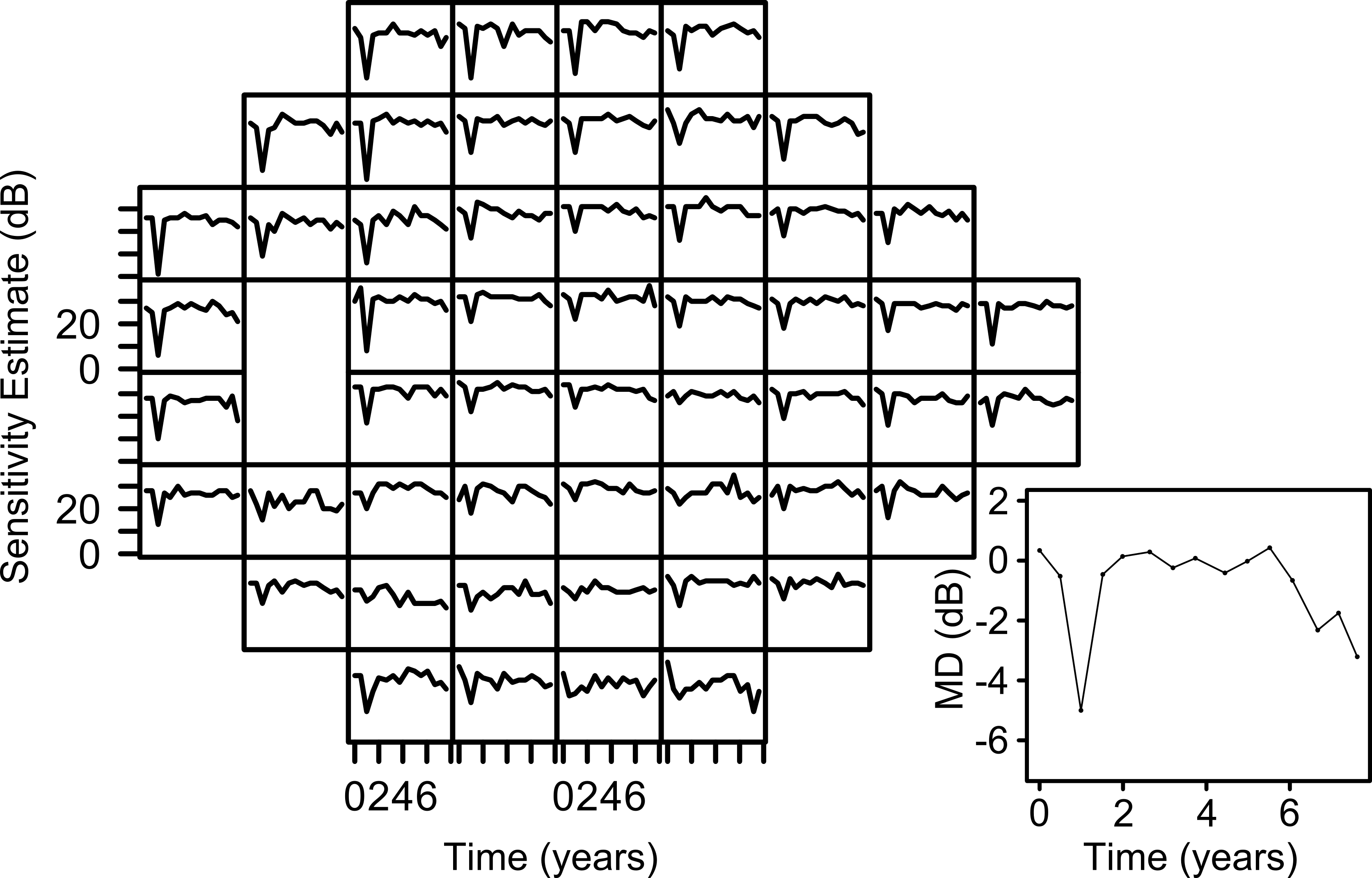}
\caption{Retinal sensitivity estimates over time for each location of the visual field in the left eye of a single glaucoma patient. A decrease in the sensitivity estimates can be seen in all locations at around 1 year. The longitudinal profile of the MD values over time are shown on the right. The visit-dependent decrease is also clear at around 1 year for the MD.}
\label{GVE_figure}
\end{center}
\end{figure}

Junoy Montolio et al. (2012) \cite{montolio_2012} explicitly modeled the global variations with known factors such as season, time of day and reliability indices. However, we speculated that other transient factors, such as fatigue, lack of concentration, or delayed reaction time may play a more important role. Since all these (as well as possibly other) factors, affect all locations belonging to the same VF, we  propose to take them together and to call them, as well as model them as the Global Visit Effects (GVEs). In this way, we can account for both the known and the unknown factors. Hence, the GVE accounts for all factors that affect all measurements of the same eye at each visit. To illustrate the importance of these factors, we show in Figure \ref{GVE_figure} the VFs over time of one eye, where all locations have a drastic decrease in sensitivity at around 1 year. From the longitudinal profiles, it is evident that this decrease is caused by something that affected all VF measurements of that visit, rather than by actual damage. To account for the visit-dependent offset at all locations, or GVE, we included a parameter, $\phi_{iet}$, in the model to capture the offset at every visit $j$ for each eye $k$ within each individual $i$. This gives,
\newline

Model 2:
\begin{eqnarray}\label{visit}
y_{iehlt} &=& \mu^{(1)}_{iehlt} + \phi_{iet} + \epsilon_{iehlt} \nonumber \\
&=& \mu^{(2)}_{iehlt} + \epsilon_{iehlt}.
 \end{eqnarray}

From an initial exploratory analysis, we observed a number of spikes in the distribution of the visit effects. To accommodate these spikes, we assumed a t-distribution for $\phi_{iet}$. The t-distribution allows greater flexibility in the distribution of random effects compared to the normal distribution, and can handle heavy tails in random effects distributions \cite{lee_2008}. Hence, we let,
\begin{eqnarray*}
\phi_{iet}   \sim  t(0, \sigma^2_\phi, 3),
\end{eqnarray*}
where $t(\mu, \sigma^2, df)$ denotes the generalized t-distribution with mean $\mu$, scale parameter $\sigma$, and $df$ degrees of freedom.

\subsection{Relationship between Variability and Sensitivity}\label{variability}

\begin{figure}[ht]
\begin{center}
\includegraphics[width=.5\textwidth]{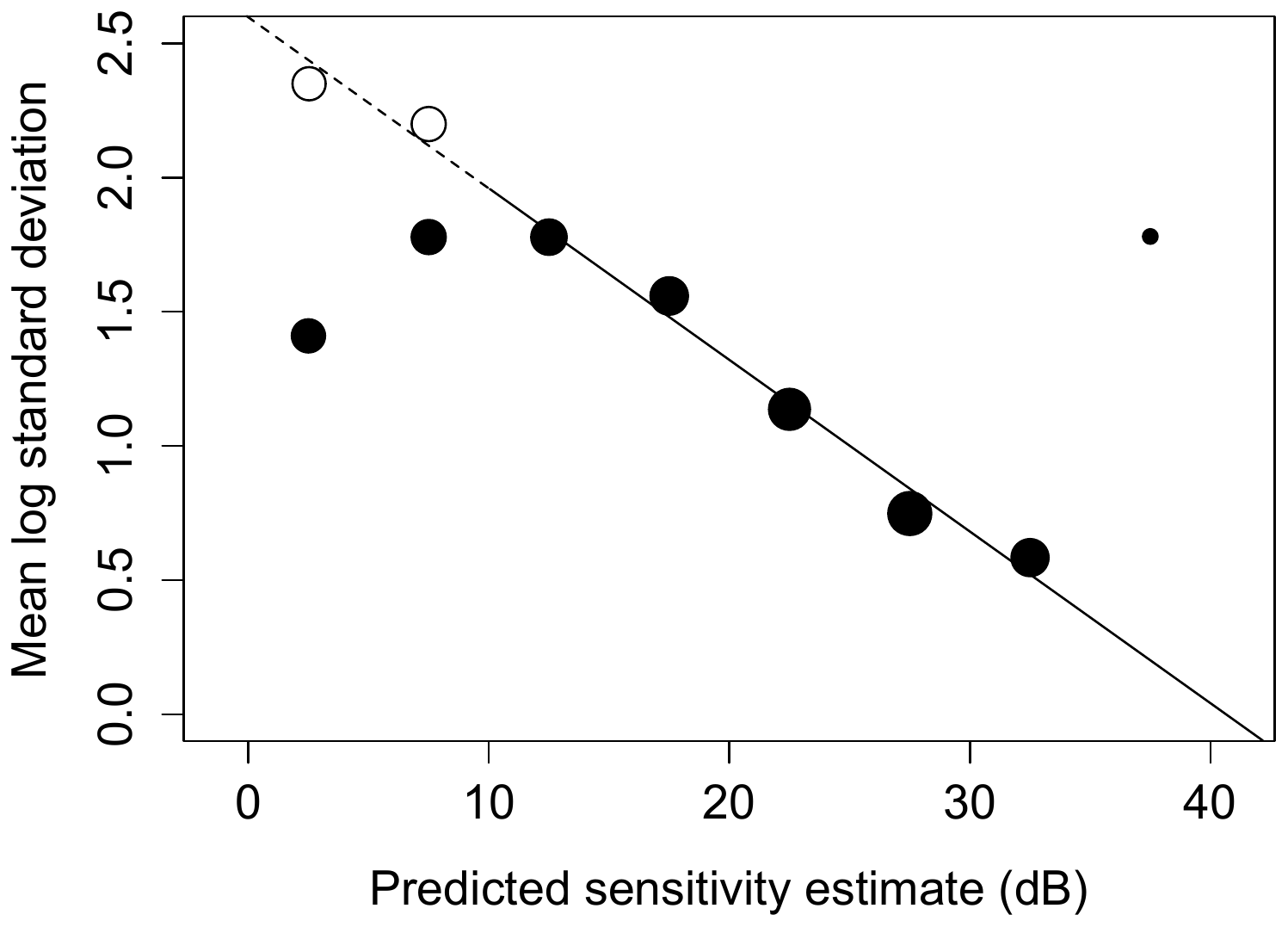}
\caption{Bubbleplot representing the mean logarithm of the standard deviation for different predicted sensitivity estimates determined using linear regression for each location. The predicted values were subdivided into groups with width 5 dB. The empty bubbles correspond to the hypothetical values, corresponding to the censored measurements, for the mean logarithm of the standard deviation for the predicted sensitivity estimates after extrapolation. The bubbles are scaled to the logarithm of the number of observations.}
\label{fig_logsigma}
\end{center}
\end{figure}

There is an association between a decline in VF sensitivity and an increase in response variability. However, values lower than 0 dB cannot be measured. This inherent censoring process introduces a positive bias at low sensitivity estimates, which is made worse by the increased variability for low sensitivity estimates. We assumed a linear relationship between the expected values of the sensitivity estimates and the logarithm of the standard deviation. However, since we were interested in modeling the latent sensitivity estimates, we extrapolated this linear relationship for predicted sensitivity estimates below 10 dB. This can be seen in Figure \ref{fig_logsigma}. In this exploratory analysis, we found that the relationship was well represented by the function $\log(SD)= A + B \times {\tt sensitivity} (dB) $, where A and B are 2.60 dB and -0.06 dB respectively. We extended Model 2 to incorporate this relationship such that,
\newline
Model 3:
\begin{eqnarray*}\label{both_eye_var}
y_{iehlt} &=&  \mu^{(2)}_{iehlt} + \epsilon_{iehlt},
 \end{eqnarray*}

and

\begin{eqnarray} \label{var_sens}
\log(\sigma_{iehlt}) &=& f\{E(y_{iehlt})\}  \nonumber \\
 &=& \beta^*_0 + \beta^*_1 \mu^{(2)}_{iehlt},
\end{eqnarray}

where $f$ is a linear function. A summary of all the parameters and their definitions for all the models is shown in Table \ref{Tab:Parameters}.
\newline

\begin{table}[!h]
\begin{center}
\tabcolsep 3 pt
\small
\caption{\small Summary of parameters included cumulatively in each of the models}
\begin{tabular}{ccl}
\hline
 Model  & Parameter & Definition\\
\hline
Model 1	& $y_{iehlt}$ & Latent sensitivity estimate  \\
		& $\beta_0$ & Population-averaged intercept  \\
		& $\beta_1$ &  Population-averaged slope \\
		& $\alpha_{0i}$ & Individual-specific intercept \\
		& $\alpha_{1i}$ & Individual-specific slope \\
		& $\gamma_{0i}$ & Eye-specific intercept \\
		& $\gamma_{1i}$ & Eye-specific slope \\
		& $\eta_{0i}$ & Hemifield-specific intercept \\
		& $\eta_{1i}$ & Hemifield-specific slope \\
		& $\lambda_{0i}$ & Location-specific intercept \\
		& $\lambda_{1i}$ & Location-specific slope \\
		& $\sigma^2$ & Variance \\
Model 2 & $\phi_{iet}$ & Global visit effect \\
		& $\sigma^2_{\phi}$ & Global Visit Effect variance \\
Model 3 & $\beta^*_0$ & Intercept in logarithm of the standard deviation \\
	    & $\beta^*_1$ & Slope in logarithm of the standard deviation \\
\hline
\\
\end{tabular}
\label{Tab:Parameters}
\end{center}
\end{table}

\section{Estimation Approach}\label{computation}

\subsection{One-stage approach}
The Bayesian approach takes into account the uncertainty in all model parameters and allows for prior information to be incorporated. Furthermore, MCMC algorithms allow greater flexibility by relaxing the strong parametric assumptions commonly used in most frequentist hierarchical models \cite{Lunn_2013,lesaffre_2012}. The classical Bayesian approach is one-stage hierarchical modeling, which has the advantage that subject-specific and overall parameters are estimated simultaneously. However, for a (relatively) large data set, this approach can be difficult or even impossible to implement for complex models with standard MCMC software. In our case, we had a total of 45,005 parameters which needed to be estimated. As a consequence, we were unable to achieve convergence in a realistic time frame and experienced computer memory limitations when using WinBUGS or JAGS. For such situations, a computationally more efficient method is needed.

\subsection{Two-stage approach}
\begin{figure}[ht]
\begin{center}
\includegraphics[width=.7\textwidth]{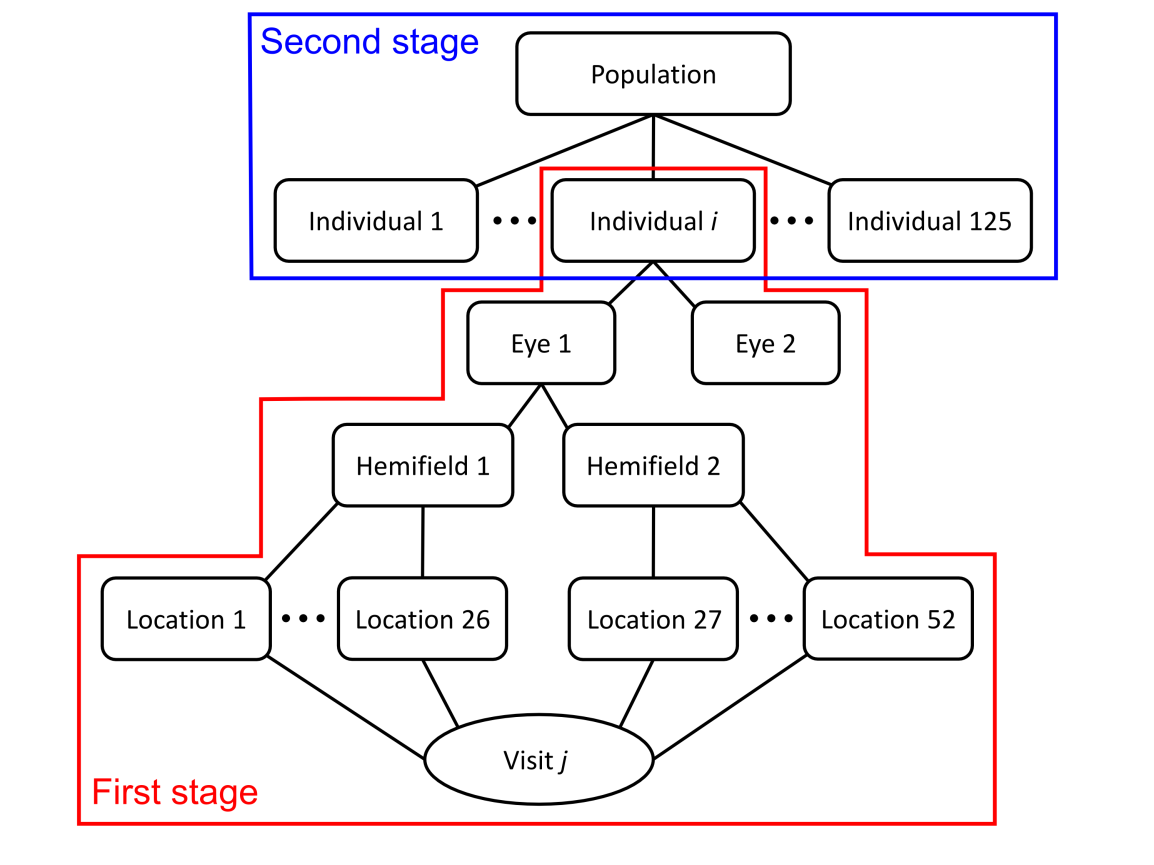}
\caption{Illustration of the hierarchical structure of the data divided into the first and second stages as done in the two-stage approach.}
\label{levels}
\end{center}
\end{figure}

Lunn et al. \cite{Lunn_2013} proposed two-stage Bayesian hierarchical modeling. The glaucoma data also exhibit an hierarchical structure, but of a more complex nature. Figure \ref{levels} illustrates the hierarchical structure of the glaucoma data, as well as the cross-classified random effects, divided into two stages. The two-stage approach allowed us to simplify the problem by splitting hierarchical models with M levels at level m*. Independent parameters of interest at level m* are obtained in stage 1 and used as proposal distributions for those parameters in stage 2. Lunn et al. illustrated this method using models with two and three levels. We applied this to a more complex model with four levels.  In our case, we split the levels at the individual level, treating each individual as their own sample. These individuals were then analyzed independently before combining them to obtain population level estimates.

\subsubsection{First stage}
In the first stage, we analyzed each individual separately. Without loss of generality, we only show this for Model 3. This becomes:

\begin{eqnarray}\label{one_stage}
y_{iehlt} &=&   \alpha_{0i} + \alpha_{1i} \mbox{\tt{years}}_{it} + \gamma_{0ie} + \gamma_{1ie} \mbox{\tt{years}}_{it}  +  \eta_{0ieh} + \eta_{1ieh} \mbox{\tt{years}}_{it}+   \nonumber \\ && \lambda_{0iehl} + \lambda_{1iehl} \mbox{\tt{years}}_{it}  + \phi_{iet} + \epsilon_{iehlt},
\end{eqnarray}

where \begin{eqnarray*}
 \gamma_{ie}  &=& \left( \begin{smallmatrix} \gamma_{0ie} \\ \gamma_{1ie} \end{smallmatrix} \right) \sim
  				  N(\left( \begin{smallmatrix} 0 \\ 0 \end{smallmatrix} \right) , \Sigma_{\gamma i} = (\begin{smallmatrix} 						  						  \Sigma_{\gamma11i} &\Sigma_{\gamma12i}  \\ \Sigma_{\gamma21i} & \Sigma_{\gamma22i} \end{smallmatrix} )); \\
 \eta_{ieh}  &=&  \left( \begin{smallmatrix} \eta_{0ieh} \\ \eta_{1ieh} \end{smallmatrix} \right) \sim
  					N (( \begin{smallmatrix} 0 \\ 0 \end{smallmatrix}) ,
  					\Sigma_{\eta i} = (\begin{smallmatrix} \Sigma_{\eta11i} &\Sigma_{\eta12i}  \\ \Sigma_{\eta21i} & \Sigma_{\eta22i} 	
  					\end{smallmatrix}) )  ; \\
 \lambda_{iehl} &=&  \left( \begin{smallmatrix} \lambda_{0iehl} \\ \lambda_{1iehl} \end{smallmatrix} \right) \sim
 					N ( \left( \begin{smallmatrix} 0 \\ 0 \end{smallmatrix} \right) ,
 					\Sigma_{\lambda i} = ( \begin{smallmatrix} \Sigma_{\lambda11i} &\Sigma_{\lambda12i}  \\ \Sigma_{\lambda21i} & \Sigma_{\lambda22i} 	
 													\end{smallmatrix})); \\
 	\phi_{iet}   &\sim&  t(0, \sigma^2_{\phi i}, 3)  \mbox{ } \mbox{and} \mbox{ }\\
 					\epsilon_{iehlt} &\sim& N(0,\sigma^2_{iehlt}),
\end{eqnarray*}

and

\begin{eqnarray*} \label{var_sens}
\log(\sigma_{iehlt}) &=&  \beta^*_{0i} + \beta^*_{1i}\mu^{(2)}_{iehlt}.
\end{eqnarray*}

One important detail about the two-stage approach is that it allows the individual variances to differ, i.e.
$\Sigma_{\gamma i}$, $\Sigma_{\eta i}$, $\Sigma_{\lambda i}$ and $\sigma^2_{\phi i}$, but also $\sigma_{iehlt}$ since the regression coefficients are now allowed to depend on the subject. This is in contrast to the one-stage model which requires the variances to be the same, i.e.
$\Sigma_{\gamma}$, $\Sigma_{\eta}$, $\Sigma_{\lambda}$ and $\sigma^2_{\phi}$. Hence, the two-stage approach is more flexible, as it can account for these differences if they are present in the data. In the Bayesian procedure prior distributions need to be assumed for all parameters. Explicit expressions for the priors are given in the Appendix. In order to prevent the second-stage sampler from becoming stuck near local posterior modes, large independent samples are needed from this first stage \cite{Lunn_2013}. To achieve this, we ran 200,000 iterations with a burn-in of 150,000 and thinning of 10, resulting in 138 samples of 5,000 iterations each for each parameter.

\subsubsection{Second stage}

In the first stage, $\alpha_i$ and $\beta^*_i$ were treated as fixed effects. These parameters need to be combined in the second stage to obtain the population-averaged effects, $\beta$ and $\beta^*$. We denote them as $\theta_i$. In the case of a meta analysis, the random effects in the first stage, which we denote as $L_i$, may not be of direct interest. Hence, these terms can be treated as nuisance parameters as done by Lunn et al. \cite{Lunn_2013}. However, for clinical applications such as ours, these may be important. In order to avoid further computational problems, we took only the covariance matrices of the random effects in the first stage to the second stage, which allowed us to re-estimate the random effects in an additional step. Since each of the elements in the matrices were treated as separate parameters in the second stage, Cholesky decomposition of the covariance matrices of the random effects, $\Sigma_{\gamma_i}$, $\Sigma_{\eta_i}$ and $\Sigma_{\lambda_i}$ respectively, was used. More specifically, for each of the above covariance matrices a full rank lower triangular matrix $U$ with real and positive diagonal entries was generated, ensuring that $UU^T$ is positive definite. We denote the Cholesky decomposition factors for all of the covariance matrices by $C_i$. Hence, we let $\left\{\theta_i, C_i \right\}$ represent parameters of interest from the first-stage. The samples of these parameters were then used as proposal distributions within a Metropolis-Hastings step in the second-stage to obtain $\left\{\theta, C \right\}$ $= \left\{\theta_i, C_i, i =1, \ldots, N \right\}$.
Three chains were initialized with different starting values for all models determined from the first-stage samples. This was done using the minimum value, the mean value and the maximum value for each parameter for every individual. Upon convergence, we computed the posterior mean, median, standard deviation  with the equal tail 95\% credible interval (CI) for all parameters of interest.

\section{Model Evaluation}\label{evaluation}

Standard approaches are applicable to the results from the first stage since this stage represents a standard analysis. Hence, we evaluated the models at this stage for each individual separately using posterior predictive checks (PPC), such as the $\chi^2$-test statistic. A further comparison of the models was done after the second stage using the Deviance Information Criterion (DIC) to determine the overall best model.

\subsection{Posterior predictive check}

We denote all parameters for individual $i$ from the first stage, i.e. $\{\theta_i, C_i, L_i\}$, as $\psi_i$. Let $\psi^1_i,\dots , \psi^K_i$ be the converged Markov chain from $p(\psi_i \mid y_{i})$. Furthermore the vector of all responses for the $i$th individual is denoted by $y_{i}$. The posterior predictive P-value (PPP) for a discrepancy measure, $D(y_{i} \mid \psi^k_i)$ is then calculated and replicated data $\tilde{y}_{i}^k$ is sampled from$p(y_{i} \mid \psi^k_i)$. $D(\tilde{y}_{i}^k,\psi^k_i)$ can then be computed for $k$ in $\{1,\dots,K\}$, and $p_{D}$ can be estimated by,

\begin{eqnarray}
\bar{p}_{D} = \frac{1}{K} \sum^K_{k=1}I[D(\tilde{y}_{i}^k, \psi^k_i) \leq D(y_{i},\psi^k_i)],
\end{eqnarray}

Here, the Gelman $\chi^2$-test statistic \cite{gelman_2013} was used as the discrepancy measure to calculate the PPC for each individual. This is defined as:

\begin{eqnarray}
 D(y_{i},\psi^k_i) = \sum^{E}_{e=1} \sum^{H}_{h=1} \sum^{L}_{l=1} \sum^{T_i}_{t=1}\frac{[y_{iehlt}-E(y_{iehlt} \mid \psi^k_i)]^2}{\mbox{\tt{var}} (y_{iehlt} \mid \psi^k_i)},
\end{eqnarray}

where $E(y_{iehlt})$ is defined as $\mu^{(2)}_{iehlt}$. A small value indicates a bad model fit. The above predictive P-values can then be contrasted against a uniform distribution to evaluate globally model fit for each individual. In general, if $\bar{p}_D$ is smaller than $0.05$ or larger than $0.95$, then this is an indication that the model might not fit the data well. Note that this procedure is, however, conservative because the data is used twice: one for model fit and one for model evaluation, see e.g. \cite{lesaffre_2012}.

\subsection{Deviance Information Criterion}

In a Bayesian framework, a common tool for model evaluation is the Deviance Information Criterion (DIC) proposed by Spiegelhalter et al. (2002) \cite{Spiegelhalter_2002}. The DIC is defined as:

\begin{eqnarray} \label{DIC}
DIC = D(\{\bar{\theta},\bar{C}\}, \bar{L}) + 2p_D = \overline{D(\{\theta,C\}, L)} + p_D,
\end{eqnarray}
\\
where
\begin{eqnarray*} \label{DIC}
p_D =  \overline{D(\{\theta,C\} L)} - D(\{\bar{\theta}, \bar{C}\}, \bar{L}).
\end{eqnarray*}
\\
In the definition of the DIC, we have the fixed effects parameters as well as the random effects which are treated as nuisance parameters in the two-stage approach. One of the disadvantages of the two-stage approach is not being able to directly obtain the random effects estimates. Since most model evaluation methods, such as the Deviance Information Criteria (DIC), require the random effect estimates for the computation, it is not clear how to evaluate models using this approach. To overcome this limitation, we propose an extension of the two-stage approach, by including an additional step based on the Method of Composition in combination with a Metropolis-within-Gibbs technique. More specifically, for the calculation of the DIC we are required to obtain a sample of the random effects from their posterior distribution $p(L_i \mid y_i)$, which is written as:



\begin{eqnarray}
p(L_i \mid y_i) = \int p(L_i \mid y_i, \Omega_i) p(\Omega_i \mid y_i) d\Omega_i,
\label{identity}
\end{eqnarray}

where
$\Omega_i = (\beta_0, \beta_1, \beta^{*}_0, \beta^{*}_1,
\alpha_{0i}, \alpha_{1i}, \Sigma_\gamma,  \Sigma_\eta,  \Sigma_\lambda,  \sigma^{2}_\phi)$
is the vector of all parameters of main interest; $\tilde{\Omega}_i$ represents the sampled values from the second stage. Identity (\ref{identity}) suggests that we can use the following simulation scheme. This is a sampling algorithm based on the Method of Composition in combination with a Metropolis-within-Gibbs technique, which is applied to each individual. For the $i$th individual the computations are done as follows:

\textit{Step 1:} For iteration $k$, let the parameters estimated in the second stage of the two-stage approach be denoted by $\tilde{\Omega}^{(k)}_i$.

\textit{Step 2:} Given  $\tilde{\Omega}^{(k)}_i$, we sample:
\begin{eqnarray*}
\gamma_{0ie}, \gamma_{1ie},
\eta_{0ieh}, \eta_{1ieh},
\lambda_{0iehl}, \lambda_{1iehl}  \mbox{  and  }
 \phi_{iet},
\end{eqnarray*}

which we denote as $L_i^{(k)}$. This is done using the Metropolis-within-Gibbs technique. Thus, for each sequence of generated parameters from the second stage, we apply MCMC sampling to obtain these estimates:
\\

\textit{Step 2A:} Initial values are determined using an optimization routine.
\\

\textit{Step 2B:} A random walk Metropolis Hastings algorithm is used for each of the levels. This is done iteratively, to take into account the correlation between the levels.
\\

\textit{Step 3:} After an inital burn-in, we save the last estimate for each parameter in $L_i^{(k)}$.
\\

\textit{Step 4:} This is repeated for K iterations, resulting in:

\begin{eqnarray*}
L_i^{(1)}, L_i^{(2)}, \dots, L_i^{(K)}
\end{eqnarray*}

Hence, in combination with the results from the second stage, we have now obtained all the parameter estimates which are needed to compute the DIC.

\section{Application to the Glaucoma study}\label{Application}

For this analysis we included both eyes from the 139 individuals belonging to the Glaucoma study. After excluding VFs with unknown reliability as indicated by the instrument, 138 individuals and 276 eyes remained. This included 4,758 VFs, resulting in a data set consisting of 14,352 VFs and 247,520 location-specific sensitivity estimates. All analyses were done taking into account censoring, and hence using the latent sensitivity values, $y_{iehlt}$.

\subsection{Results}

\begin{figure}[!ht]
 \hspace*{\fill}%
    \subfloat{%
      \includegraphics[width=0.3\textwidth]{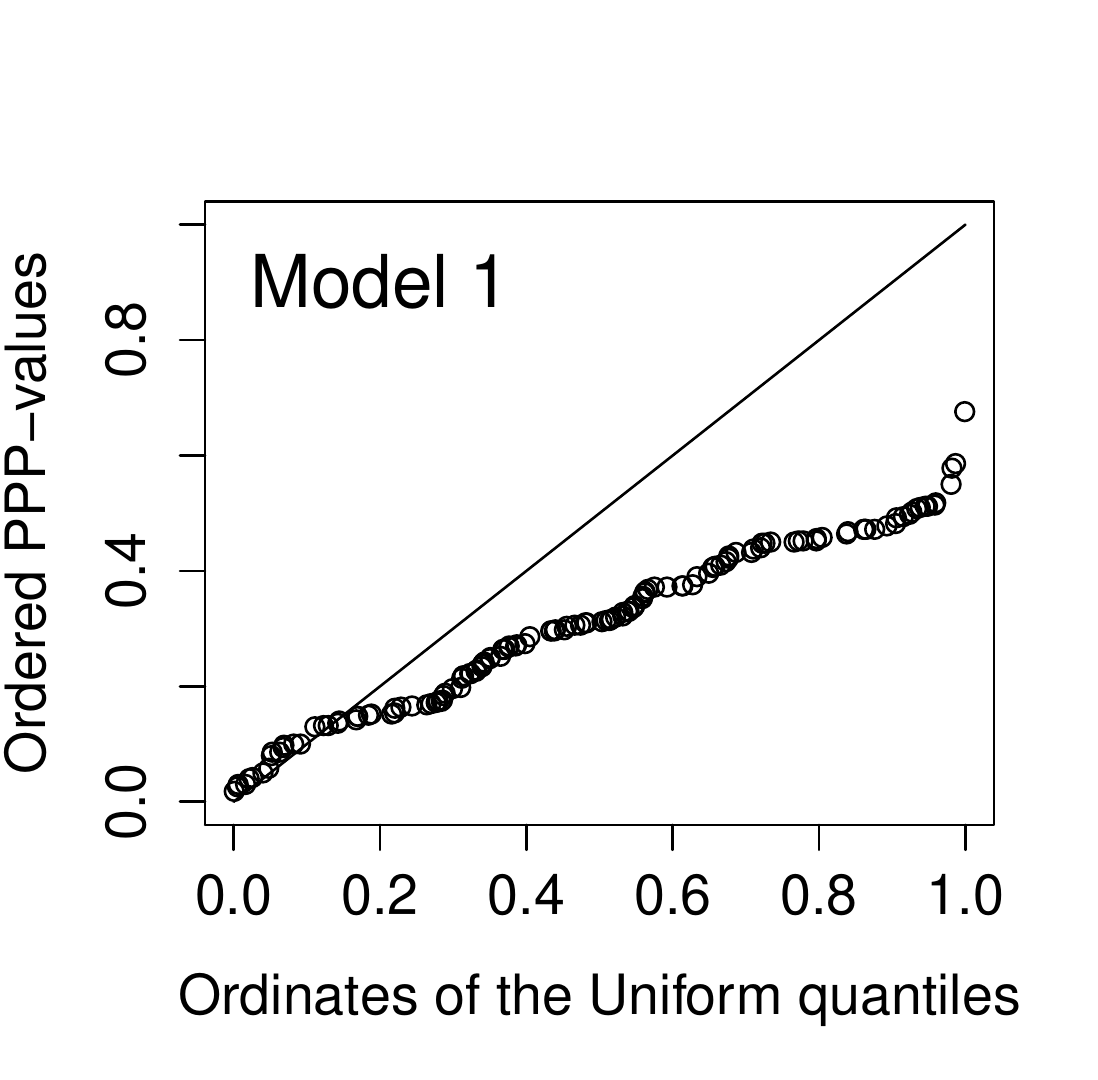}
    }
    \hfill
    \subfloat{%
      \includegraphics[width=0.3\textwidth]{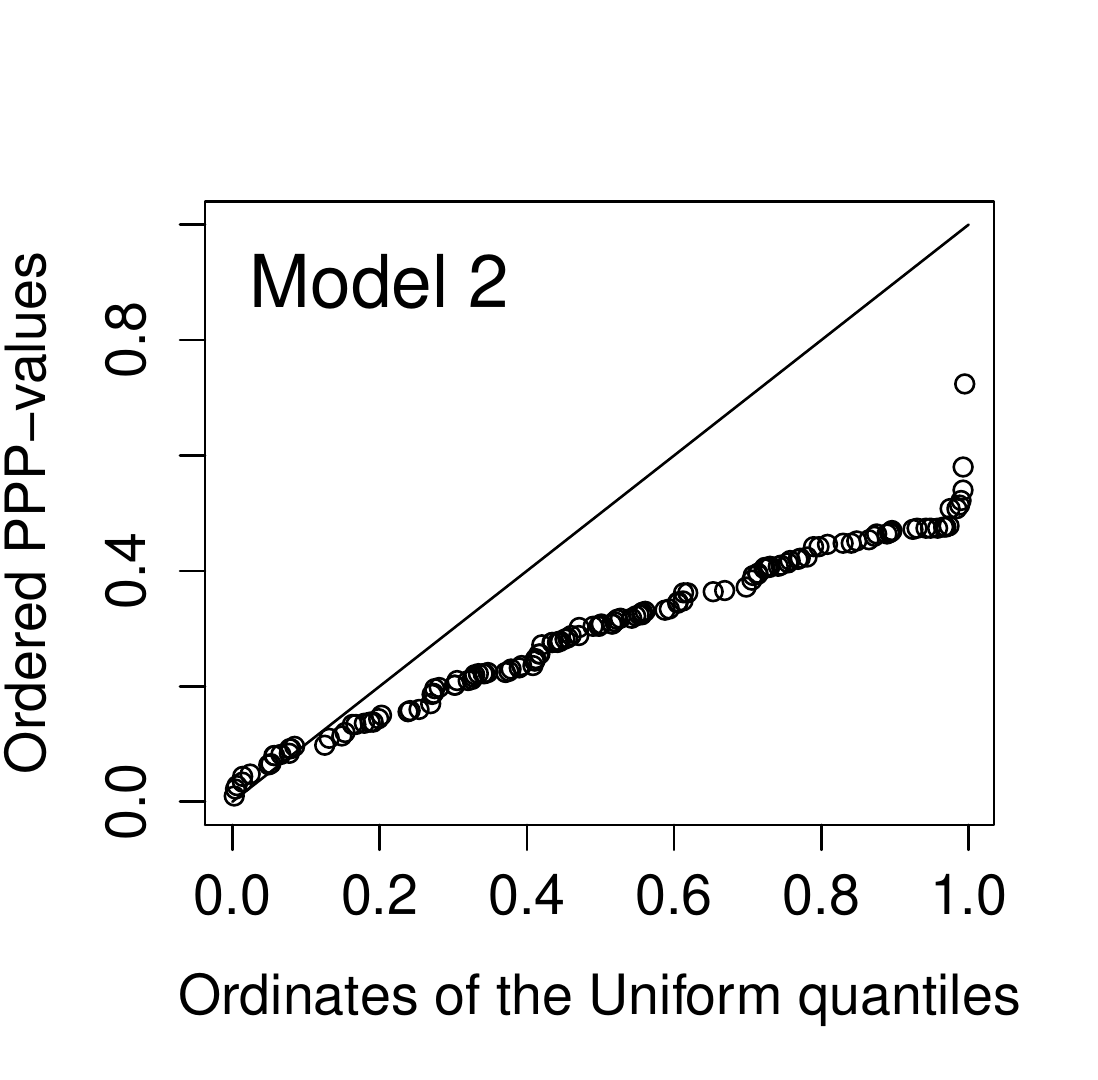}
    }
  \hfill
    \subfloat{%
      \includegraphics[width=0.3\textwidth]{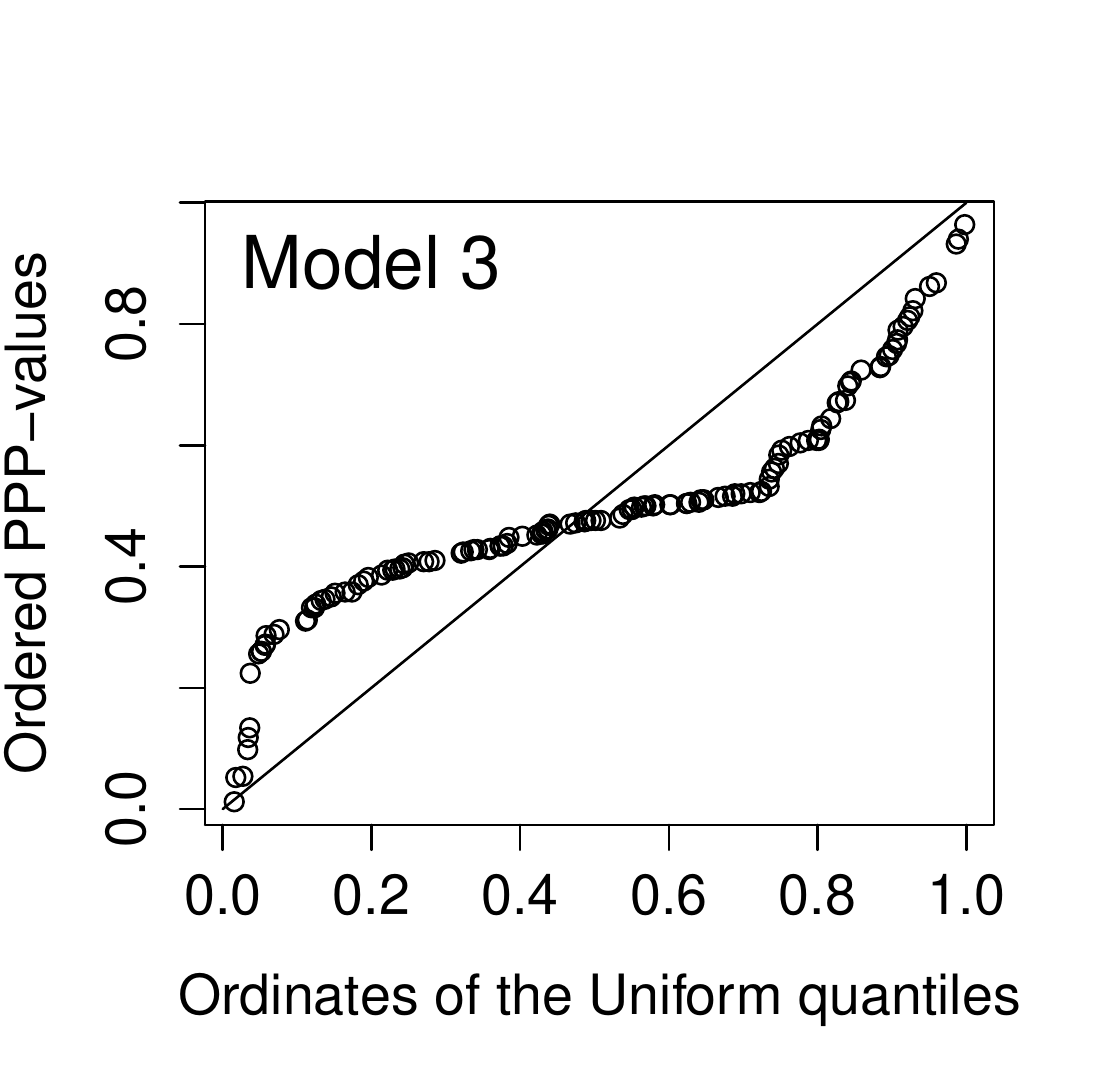}
    }
    \caption{Posterior predictive check for each of the models across all individuals}
    \label{ppp}
 \hspace*{\fill}%
  \end{figure}

The two-stage approach is advantageous, as it allows us to do exploratory analyses at the individual level in order to simplify the model before combining the samples in the second stage. In order to compare the models, we can evaluate the outcome of the PPC using graphical output. The PPP-values denoted by $\bar{p}_D$ were computed for every individual. Figure \ref{ppp} shows the ordered PPP-values for each of the models. Model 1 has a mean $\bar{p}_D=0.30$, Model 2 a mean $\bar{p}_D=0.30$ and Model 3 a mean $\bar{p}_D=0.50$. From this, it appears that Model 3 has the best fit. This approach gives a good indication of whether the models fit the data, specifically for each individual.

\begin{figure}[ht]
\begin{center}
\includegraphics[width=.6\textwidth]{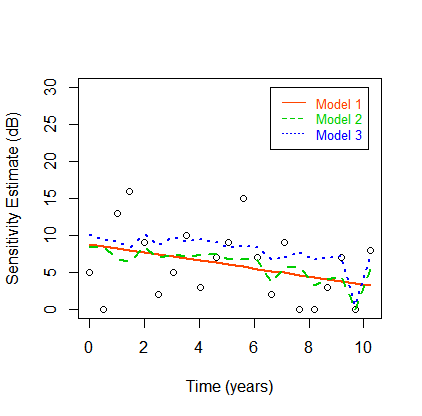}
\caption{Scatter plot representing the retinal sensitivity estimates over time for 1 location of the VF. The lines represent the model fits for each of the 3 models}
\label{one_location}
\end{center}
\end{figure}

An example of the model fits for 1 location is shown in Figure \ref{one_location}. The posterior summary statistics from the second stage are listed in Table \ref{Tab:two-stage-results} for each of the models. A difference in DIC of more than 10 indicates that the model with the lowest DIC has a better fit. Using the DIC to compare the models, Model 2 (DIC=$-9.045 e^{+07}$) performed better than Model 1 (DIC = $-8.827e^{+07}$), with Model 3 (DIC = $-1.792e^{+26}$) performing the best overall. Using the results from Model 3, the population intercept ($\beta_0$) was 19.82 dB with an average slope ($\beta_1$) of  -0.31 dB per year. The intercept ($\beta^*_0$) and slope ($\beta^*_1$) for the logarithm of the standard deviation was 2.82 dB and -0.08 dB respectively. This corresponds to the 2.60 dB and -0.06 dB which was found in the exploratory analysis shown in Figure \ref{fig_logsigma}.

\begin{table}[!h]
\begin{center}
\tabcolsep 3 pt
\small
\caption{\small Posterior summary statistics for the three models using the two-stage approach}
\begin{tabular}{rcccccccccccr}
\hline
 Model  & 1 & &  &  2 &  &  & 3 & & & \\
\hline
Parameter &	 mean & sd & 95\% CI
		  &	 mean & sd & 95\% CI
		  &	 mean & sd & 95\% CI  \\
\hline
$\beta_{0}$ & 18.95 & 0.72 & (17.53 ; 20.36)	
& 20.42 & 0.73 & (18.95 ; 21.84)
& 19.89 & 0.77 & (18.36 ; 21.37)  \\
$\beta_{1}$ & -0.22 & 0.0 & (-0.33 ; -0.13)
& -0.21 & 0.05 & (-0.31 ; -0.11)
& -0.31 & 0.05 & (-0.41 ; 0.20) \\
$\beta^*_{0}$ &  &  &
&  &  &
& 2.82 & 0.06 & (2.70 ; 2.95)  \\
$\beta^*_{1}$ &  &  &
& &  &
& -0.08 & 0.08 & (-0.08 ; -0.07) \\
$\sigma^2$  &  13.42 & 0.66 & (12.17 ; 14.75)
& 11.51 & 0.56 & (10.45 ; 12.68)	
&   	  \\
$\sigma^2_\phi$  &  &  &
& 0.62 & 0.05 & (0.52 ; 0.73)
& 1.87 & 0.04 & (1.81 ; 1.96)  	  \\
\hline
DIC &  & & $-8.827e^{+07}$ &  & &  $-9.045e^{+07}$ &&&  $-1.792e^{+26}$  \\
\hline
\\
\end{tabular}
\label{Tab:two-stage-results}
\end{center}
\end{table}

\subsection{Clinical Implications}

With the GVE, we account for those factors, as well as those which can not be measured such as fatigue and delayed reaction time. Including the GVE showed a significant improvement in the model fit. Hence, by taking into account the GVE we were able to take into account a large part of the variability and obtain better estimates of the true rate of progression. By including the relationship between variability and sensitivity shows a further improvement in the model fit. The function which describes this relationship was consistent to that found by Henson et al \cite{Henson_2000}, however it was not shown previously how to include this relationship in a model, or whether including it would improve the model fit. By including both of these aspects, we were able to improve the estimation of the true underlying progression and determine the real evolution of the sensitivity over time. Hence, these improved estimates could aid clinicians in optimizing treatment strategies.

\section{Discussion}\label{Discussion}

In this paper, we proposed a method to model point-wise VFs taking into account the complexity of psychophysical testing of visual function in glaucoma. The model is advantageous in dealing with the high measurement variability, and could be extended for the prediction of future VFs. Although it was possible to use the one-stage approach with simplied versions of the model or with smaller datasets, it was not possible to perform these analysis on the full data with a complex model as it was with the two-stage approach. The two-stage approach can be implemented in standard MCMC software. The relevant computations for the first-stage can be carried out in JAGS \cite{Plummer_2003}, WinBUGS or OpenBUGS \cite{Lunn_2009} software and the second-stage using OpenBUGS software \cite{Lunn_2013}. However, for the second stage an add-on program is needed. For more details on setting up OpenBUGS for performing the two-stage analyses, we refer to \cite{Lunn_2013}. More information regarding the computations done in this paper can be obtained by emailing the first author. These computations can be easily tuned to adapt to other data sets by any practitioner. 

The two-stage method is advantageous as it allows us to do exploratory analysis at an individual level. Hence, we are able to simplify and improve the model before combining it at a population level. Limited simulations showed that the one- and two-stage approaches gave similar results if the variances were the same for all individuals. The two-stage approach assumes a more flexible method. However, there is the additional difficulty in constraining the parameters across individuals. One disadvantage of this approach is that it does not provide the required components to evaluate the fit and predictive ability of the model using the Deviance Information Criterion (DIC). In order to calculate the DIC and compare different competing models for our data fitted using the two-stage approach, we suggested a Monte Carlo scheme based on a Metropolis-within-Gibbs algorithm.

Other issues, which we see as future research directions, is to look at the optimal choice of the level where the data should be split. Extensions include exploiting the spatial nature of the data and capitalizing on the specific spatial organization of the nerve fibres in the eye \cite{Erler_2014}.

\newpage
\bibliographystyle{plain}	

	\label{References}	

\newpage
\appendix
\numberwithin{equation}{section}
\section{Appendix}\label{appendix}

\subsection{One Stage Approach}
\subsubsection{Full Model}

\begin{eqnarray}\label{one_stage}
y_{iehlt} &=& \beta_0 + \beta_1 \mbox{\tt{years}}_{it}  +  \alpha_{0i} + \alpha_{1i}\mbox{\tt{years}}_{it} + \gamma_{0ie} + \gamma_{1ie}\mbox{\tt{years}}_{it}  + \nonumber \\ &&  \eta_{0ieh} + \eta_{1ieh}\mbox{\tt{years}}_{it}+  \lambda_{0iehl} + \lambda_{1iehl}\mbox{\tt{years}}_{ij}  + \phi_{iet} + \epsilon_{iehlt}
\end{eqnarray}

where $\epsilon_{iehlt} \sim N(0,\sigma^2_{iehlt})$ and $log(\sigma_{iehlt}) = \beta^*_0 + \beta^*_1 \mu^*_{iehlt}$.

\subsubsection{Priors}
In the Bayesian procedure prior distributions need to be stipulated for all parameters. When no prior information is available then the prior distribution should reflect this. In this case a vague prior is a natural choice.

\begin{eqnarray*}
 \beta_b  &\sim&   N(0, 10^8)  \mbox{ for } b=0,1;  \\
  \beta^*_q  &\sim&   N(0, 10^8)  \mbox{ for } q = 0, 1  \\
     \alpha_{i}  &=&  \left( \begin{smallmatrix} \alpha_{0i} \\ \alpha_{1i} \end{smallmatrix} \right) \sim
  				  N(\left( \begin{smallmatrix} 0 \\ 0 \end{smallmatrix} \right) ,
  				  \Sigma_\alpha = \left(\begin{smallmatrix} \Sigma_{\alpha11} &\Sigma_{\alpha12}  \\ \Sigma_{\alpha21} & \Sigma_{\alpha22} 		  							  \end{smallmatrix} \right) ) ; \\
 \gamma_{ie}  &=& \left( \begin{smallmatrix} \gamma_{0ie} \\ \gamma_{1ie} \end{smallmatrix} \right) \sim
  				  N(\left( \begin{smallmatrix} 0 \\ 0 \end{smallmatrix} \right) , \Sigma_\gamma = (\begin{smallmatrix} 						  						  \Sigma_{\gamma11} &\Sigma_{\gamma12}  \\ \Sigma_{\gamma21} & \Sigma_{\gamma22} \end{smallmatrix} )); \\
 \eta_{ieh}  &=&  \left( \begin{smallmatrix} \eta_{0ieh} \\ \eta_{1ieh} \end{smallmatrix} \right) \sim
  					N (( \begin{smallmatrix} 0 \\ 0 \end{smallmatrix}) ,
  					\Sigma_\eta = (\begin{smallmatrix} \Sigma_{\eta11} &\Sigma_{\eta12}  \\ \Sigma_{\eta21} & \Sigma_{\eta22} 	
  					\end{smallmatrix}) )  ; \\
 \lambda_{iehl} &=&  \left( \begin{smallmatrix} \lambda_{0iehl} \\ \lambda_{1iehl} \end{smallmatrix} \right) \sim
 					N ( \left( \begin{smallmatrix} 0 \\ 0 \end{smallmatrix} \right) ,
 					\Sigma_\lambda = ( \begin{smallmatrix} \Sigma_{\lambda11} &\Sigma_{\lambda12}  \\ \Sigma_{\lambda21} & \Sigma_{\lambda22} 	
 													\end{smallmatrix})) ; \\											\phi_{iet}   &\sim &  t(0, \sigma^2_\phi, 3)  \mbox{ } \mbox{and} \mbox{ }\\
 					\epsilon_{iehlt} &\sim& N(0,\sigma^2) .
\end{eqnarray*}

The variance was given a vague inverse gamma prior. The covariance matrices of the random effects, i.e. $\Sigma_{\gamma}$, $\Sigma_{\eta}$ and $\Sigma_{\lambda}$, were given a vague inverse Wishart distribution with degrees of freedom equal to the number of dimensions and small diagonal values for the scale matrix.

\subsection{Two Stage Approach}
\subsubsection{Full Model}

\begin{eqnarray}\label{one_stage}
y_{iehlt} &=& \beta_0 + \beta_1\mbox{\tt{years}}_{it} +  \alpha_{0i} + \alpha_{1i}\mbox{\tt{years}}_{it} + \gamma_{0ie} + \gamma_{1ie}\mbox{\tt{years}}_{it}  + \nonumber \\ &&  \eta_{0ieh} + \eta_{1ieh}\mbox{\tt{years}}_{it}+  \lambda_{0iehl} + \lambda_{1iehl}\mbox{\tt{years}}_{it}  + \phi_{iet} + \epsilon_{iehlt}
\end{eqnarray}

where $\epsilon_{iehlt} \sim N(0,\sigma^2_{iehlt})$ and $log(\sigma_{iehlt}) = \beta^*_0 + \beta^*_1 \mu^*_{iehlt}$.

\subsubsection{Priors}

\begin{eqnarray*}
 	\phi_{iet}   &\sim&  t(0, \sigma^2_{\phi i}, 3) ; \sigma^2_{\phi_i} \sim N(\sigma^2_\phi, \Sigma_\phi) \\
  \alpha_{gi} &\sim&   N(0, 10^8)  \mbox{ for } g=0,1; \\
 \gamma_{ie}  &=& \left( \begin{smallmatrix} \gamma_{0ie} \\ \gamma_{1ie} \end{smallmatrix} \right) \sim
  				  N(\left( \begin{smallmatrix} 0 \\ 0 \end{smallmatrix} \right) , \Sigma_{\gamma i} = (\begin{smallmatrix} 						  						  \Sigma_{\gamma11i} &\Sigma_{\gamma12i}  \\ \Sigma_{\gamma21i} & \Sigma_{\gamma22i} \end{smallmatrix} )); \\
 \eta_{ieh}  &=&  \left( \begin{smallmatrix} \eta_{0ieh} \\ \eta_{1ieh} \end{smallmatrix} \right) \sim
  					N (( \begin{smallmatrix} 0 \\ 0 \end{smallmatrix}) ,
  					\Sigma_{\eta i} = (\begin{smallmatrix} \Sigma_{\eta11i} &\Sigma_{\eta12i}  \\ \Sigma_{\eta21i} & \Sigma_{\eta22i} 	
  					\end{smallmatrix}) )  ; \\
 \lambda_{iehl} &=&  \left( \begin{smallmatrix} \lambda_{0iehl} \\ \lambda_{1iehl} \end{smallmatrix} \right) \sim
 					N ( \left( \begin{smallmatrix} 0 \\ 0 \end{smallmatrix} \right) ,
 					\Sigma_{\lambda i} = ( \begin{smallmatrix} \Sigma_{\lambda11i} &\Sigma_{\lambda12i}  \\ \Sigma_{\lambda21i} & \Sigma_{\lambda22i} 	
 													\end{smallmatrix}))
\mbox{ } \mbox{and} \mbox{ }\\
 					\epsilon_{iehlt} &\sim& N(0,\sigma^2_{iehlt}) .
\end{eqnarray*}

Using Cholesky decompostion, $\Sigma_{\gamma_i}$, $\Sigma_{\eta_i}$ and $\Sigma_{\lambda_i}$ become,
 \begin{eqnarray*}
 C_{r\gamma_i} \sim N(C_{r\gamma}, 10^8)  \mbox{ for } r= 1,2,3  \\
 C_{r\eta_i} \sim N(C_{r\eta}, 10^8)  \mbox{ for } r= 1,2,3  \\
 C_{r\lambda_i} \sim N(C_{r\lambda}, 10^8)  \mbox{ for } r= 1,2,3
  \end{eqnarray*}

\subsubsection{First Stage Model}

\begin{eqnarray}\label{one_stage}
y_{iehlt} &=&   \alpha_{0i} + \alpha_{1i} \mbox{\tt{years}}_{it} + \gamma_{0ie} + \gamma_{1ie} \mbox{\tt{years}}_{it}  +  \eta_{0ieh} + \eta_{1ieh} \mbox{\tt{years}}_{it}+   \nonumber \\ && \lambda_{0iehl} + \lambda_{1iehl} \mbox{\tt{years}}_{it}  + \phi_{iet} +  \epsilon_{iehlt}
\end{eqnarray}

where $\epsilon_{iehlt} \sim N(0,\sigma^2_{iehlt})$ and $log(\sigma_{iehlt}) = \beta^*_{0i} + \beta^*_{1i} \mu^*_{iehlt}$.

\subsubsection{First Stage Priors}

\begin{eqnarray*}
 \beta^*_{qi}  &\sim&  N(0, 10^8)  \mbox{ for } q = 0, 1  \\
 \alpha_{gi}  &\sim&   N(0, 10^8)  \mbox{ for } g= 0,1  \\
  \phi_{iet}   &\sim&  t(0, \sigma^2_{\phi_i}, 3)\\
 \gamma_{ie}  &=&  (\gamma_{0ie}, \gamma_{1ie})^T \sim  N_2((0,0)^T, \Sigma_{\gamma_i})\\
\eta_{ieh}  &=&  (\eta_{0ieh}, \eta_{1ieh})^T \sim  N_2((0,0)^T, \Sigma_{\eta_i}) \\
\lambda_{iehl} &=& (\lambda_{0iehl}, \lambda_{1iehl})^T \sim N_2((0,0)^T,  \Sigma_{\lambda_i}).
\end{eqnarray*}

The variance was given a vague inverse gamma prior. The covariance matrices of the random effects, i.e. $\Sigma_{\gamma_i}$, $\Sigma_{\eta_i}$ and $\Sigma_{\lambda_i}$, were given a vague inverse Wishart distribution with degrees of freedom equal to the number of dimensions and small diagonal values for the scale matrix.

\subsubsection{Second Stage Priors}

\begin{eqnarray*}
 \beta^*_{qi}  &\sim&   N(\beta^*_q, 10^8)  \mbox{ for } q = 0,1  \\
 \alpha_{i}  &=&  (\alpha_{0i}, \alpha_{1i})^T \sim  N_2((\beta_0,\beta_1)^T, \Sigma_{\alpha}) \\
\sigma^2_{\phi_i} &\sim& N(\sigma^2_\phi, \Sigma_\phi)\\
  C_{r\gamma_i} &\sim& N(C_{r\gamma}, 10^8)  \mbox{ for } r = 1,2,3  \\
 C_{r\eta_i} &\sim& N(C_{r\eta}, 10^8)  \mbox{ for } r = 1,2,3  \\
 C_{r\lambda_i} &\sim& N(C_{r\lambda}, 10^8)  \mbox{ for } r = 1,2,3
\end{eqnarray*}

\section{Methods}

Let,
\begin{eqnarray*}
\beta^* &=& \{(\beta^*_{0i}, \beta^*_{1i}), \mbox{ for } i = 1,\dots,N \}  \\
\alpha &=& \{(\alpha_{0i}, \alpha_{1i}), \mbox{ for } i = 1,\dots,N \} \\
C_{\gamma} &=& \{(C_{1\gamma_i}, C_{2\gamma_i}, C_{3\gamma_i}), \mbox{ for } i = 1,\dots,N \} \\
C_{\eta} &=& \{(C_{1\eta_i}, C_{2\eta_i}, ... , C_{3\eta_i}), \mbox{ for } i = 1,\dots,N \} \\
C_{\lambda} &=& \{(C_{1\lambda_i}, C_{2\lambda_i}, ... , C_{3\lambda_i}), \mbox{ for } i = 1,\dots,N \}
\end{eqnarray*}
\\
Then, from Lunn et al. (2013),  
$\theta$ $=$ parameters of interest:  $\alpha,  \beta^*, C_{\gamma}, C_{\eta}, C_{\lambda}$ \\
$\lambda$ $=$ nuisance parameters: $\phi, \gamma, \eta, \lambda$ \\
$\mu$ $=$ mean of $\theta$: $\beta^{(1)}, \beta^{(2)}, \bar{C_{\gamma}}, \bar{C_{\eta}}, \bar{C_{\lambda}}$ \\
$\Sigma$ $=$ covariance matrix of $\theta$: $ \Sigma_{\beta(1)}, \Sigma_{\beta(2)},  \Sigma_{C_{\gamma}}, \Sigma_{C_{\eta}}, \Sigma_{C_{\lambda}}$

\subsection{Full Model}

The joint posterior distribution is given by,
\begin{eqnarray}\label{Full}
 && p(\beta^{(1)}, \beta^{(2)}, \bar{C_{\gamma}}, \bar{C_{\eta}}, \bar{C_{\lambda}},\Sigma_{\beta(1)}, \Sigma_{\beta(2)}, \Sigma_{C_{\gamma}}, \Sigma_{C_{\eta}}, \Sigma_{C_{\lambda}},
 \alpha, \beta^*,    C_{\gamma}, C_{\eta}, C_{\lambda}, \phi, \gamma, \eta, \lambda| y) \nonumber \\  &&
\propto
p(\beta^{(1)})p(\beta^{(2)})p(C_{\gamma})p(C_{\eta})p(C_{\lambda})
p(\bar{C_{\gamma}})p(\bar{C_{\eta}})p(\bar{C_{\lambda}})   p( \Sigma_{\beta(1)})p(\Sigma_{\beta(2)})  p(\Sigma_{C_{\gamma}})p(\Sigma_{C_{\eta}})p(\Sigma_{C_{\lambda}})   \times \nonumber \\ &&
 \prod_{i=1}^{N}\left\{ p(y_i|\alpha_i, \beta^*_{i}, C_{\gamma_i}, C_{\eta_i}, C_{\lambda_i},\phi_i, \gamma_i, \eta_i, \lambda_i)
p(\alpha_i| \beta^{(1)},\Sigma_{\beta(1)} ) \right.\left.p( \beta^*_{i}|\beta^{(2)},\Sigma_{\beta(2)}) \right.  \times  \nonumber \\ &&
\left. p(C_{\gamma_i}|\bar{C_{\gamma}}, \Sigma_{C_{\gamma}})
p(C_{\eta_i}|\bar{C_{\eta}},\Sigma_{C_{\eta}})  p(C_{\lambda_i}|  \bar{C_{\lambda}},  \Sigma_{C_{\lambda}})p(\phi, \gamma, \eta, \lambda)   \right\}
\end{eqnarray}

\subsection{First Stage}
We analyse all individuals independently from the joint posterior distribution of each $\alpha_i,  \beta^*_{i}, C_{\gamma_i}, C_{\eta_i}, C_{\lambda_i}$  conditional on $y_i$ alone,

\begin{eqnarray}
 p( \alpha_i,\beta^*_{i},    C_{\gamma_i}, C_{\eta_i}, C_{\lambda_i}, \phi_i, \gamma_i, \eta_i, \lambda_i| y_i)  \propto && p(y_i | \alpha_i, \beta^*_{i},    C_{\gamma_i}, C_{\eta_i}, C_{\lambda_i}, \phi_i, \gamma_i, \eta_i, \lambda_i) \times
\nonumber \\ && p(\alpha_i,  \beta^*_{i},    C_{\gamma_i}, C_{\eta_i}, C_{\lambda_i})   p(\phi_i, \gamma_i, \eta_i, \lambda_i)
\end{eqnarray}

\newpage
\subsection{Second Stage}
From the distributions in (2.1) these are given by,

\begin{eqnarray}
 &&p(\beta^{(1)}, \beta^{(2)}, \bar{C_{\gamma}}, \bar{C_{\eta}}, \bar{C_{\lambda}}|\Sigma_{\beta(1)}, \Sigma_{\beta(2)}, \Sigma_{C_{\gamma}}, \Sigma_{C_{\eta}}, \Sigma_{C_{\lambda}},\alpha, \beta^*, C_{\gamma}, C_{\eta}, C_{\lambda},\phi, \gamma, \eta, \lambda,y) \nonumber \\
 && \propto
p(\beta^{(1)}, \beta^{(2)}, \bar{C_{\gamma}}, \bar{C_{\eta}}, \bar{C_{\lambda}})
\prod_{i=1}^{N}p(\alpha_i, \beta^*_{i}, C_{\gamma_i}, C_{\eta_i}, C_{\lambda_i}|
\beta^{(1)}, \beta^{(2)},  \bar{C_{\gamma}}, \bar{C_{\eta}}, \bar{C_{\lambda}}, \nonumber \\ &&
\Sigma_{\beta(1)}, \Sigma_{\beta(2)},  \Sigma_{C_{\gamma}}, \Sigma_{C_{\eta}}, \Sigma_{C_{\lambda}})
\\
\nonumber\\
\nonumber\\
&&p(\Sigma_{\beta(1)}, \Sigma_{\beta(2)},  \Sigma_{C_{\gamma}}, \Sigma_{C_{\eta}}, \Sigma_{C_{\lambda}}|\beta^{(1)}, \beta^{(2)},  \bar{C_{\gamma}}, \bar{C_{\eta}}, \bar{C_{\lambda}},\alpha,  \beta^*_, C_{\gamma}, C_{\eta}, C_{\lambda},\phi, \gamma, \eta, \lambda,y) \nonumber \\ && \propto
p( \Sigma_{\beta(1)}, \Sigma_{\beta(2)}, \Sigma_{C_{\gamma}}, \Sigma_{C_{\eta}}, \Sigma_{C_{\lambda}})
\prod_{i=1}^{N} p(\alpha_i, \beta^*_{i}, C_{\gamma_i}, C_{\eta_i}, C_{\lambda_i}|\beta^{(1)}, \beta^{(2)}, \bar{C_{\gamma}}, \bar{C_{\eta}}, \bar{C_{\lambda}},
\nonumber \\ &&
\Sigma_{\beta(1)}, \Sigma_{\beta(2)}, \Sigma_{C_{\gamma}}, \Sigma_{C_{\eta}}, \Sigma_{C_{\lambda}})
\\
\nonumber \\ \nonumber \\
&&p(\alpha_i,  \beta^*_{i}, C_{\gamma_i}, C_{\eta_i}, C_{\lambda_i},\phi_i, \gamma_i, \eta_i, \lambda_i|\beta^{(1)}, \beta^{(2)}, \bar{C_{\gamma}}, \bar{C_{\eta}}, \bar{C_{\lambda}},\Sigma_{\beta(1)}, \Sigma_{\beta(2)}, \Sigma_{C_{\gamma}}, \Sigma_{C_{\eta}}, \Sigma_{C_{\lambda}},y)
\nonumber \\ && \propto
p(y_i|\alpha_i, \beta^*_{i}, C_{\gamma_i}, C_{\eta_i}, C_{\lambda_i},\phi_i, \gamma_i, \eta_i, \lambda_i,\phi_i, \gamma_i, \eta_i, \lambda_i)p(\alpha_i, \beta^*_{i}, C_{\gamma_i}, C_{\eta_i}, C_{\lambda_i}|\beta^{(1)}, \beta^{(2)},
\nonumber \\ &&
 \bar{C_{\gamma}}, \bar{C_{\eta}}, \bar{C_{\lambda}},\Sigma_{\beta(1)}, \Sigma_{\beta(2)},
 \Sigma_{C_{\gamma}}, \Sigma_{C_{\eta}}, \Sigma_{C_{\lambda}})p(\phi_i, \gamma_i, \eta_i, \lambda_i)
\nonumber \\ && i= 1, ... , N.
\end{eqnarray}
\newline
The distributions from (B.3) and (B.4) are available in closed form and can hence we can sample from them directly by
using standard algorithms. For the distributions (B.5) we use the distributions in (B.2) as the proposal distributions within
a Metropolis-Hastings step. For this, the target-to-proposal ratio can be simplified to,

\begin{eqnarray}
&& C(\alpha_i, \beta^*_{i}, C_{\gamma_i}, C_{\eta_i}, C_{\lambda_i},\phi_i, \gamma_i, \eta_i, \lambda_i) \nonumber\\
& = &\frac{p(y_i|\alpha_i,  \beta^*_{i}, C_{\gamma_i}, C_{\eta_i}, C_{\lambda_i},\phi_i, \gamma_i, \eta_i, \lambda_i)}
{p(y_i|\alpha_i, \beta^*_{i}, C_{\gamma_i}, C_{\eta_i}, C_{\lambda_i},\phi_i, \gamma_i, \eta_i, \lambda_i}) \times
\nonumber\\
&& \frac{p(\alpha_i, \beta^*_{i}, C_{\gamma_i}, C_{\eta_i}, C_{\lambda_i}|\beta^{(1)}, \beta^{(2)},
\bar{C_{\gamma}}, \bar{C_{\eta}}, \bar{C_{\lambda}},\Sigma_{\beta(1)}, \Sigma_{\beta(2)},
 \Sigma_{C_{\gamma}}, \Sigma_{C_{\eta}}, \Sigma_{C_{\lambda}})p(\phi_i, \gamma_i, \eta_i, \lambda_i)}
{p(\alpha_i, \beta^*_{i}, C_{\gamma_i}, C_{\eta_i}, C_{\lambda_i})p(\phi_i, \gamma_i, \eta_i, \lambda_i)} \nonumber \\
& = & \frac{p(\alpha_i,  \beta^*_{i}, C_{\gamma_i}, C_{\eta_i}, C_{\lambda_i}|\beta^{(1)}, \beta^{(2)}, \bar{C_{\gamma}}, \bar{C_{\eta}}, \bar{C_{\lambda}}, \Sigma_{\beta(1)}, \Sigma_{\beta(2)}, \Sigma_{C_{\gamma}}, \Sigma_{C_{\eta}}, \Sigma_{C_{\lambda}})}{p(\alpha_i, \beta^*_{i}, C_{\gamma_i}, C_{\eta_i}, C_{\lambda_i})}
\end{eqnarray}

\end{document}